\documentclass[aps,prd,superscriptaddress,groupedaddress,nofootinbib,nobibnotes]{revtex4}

\usepackage{graphicx}
\usepackage{dcolumn}
\usepackage{bm}
\usepackage{amssymb}
\usepackage{epstopdf}
\usepackage{amsmath}
\usepackage{amsfonts}
\usepackage{color}
\usepackage{mathrsfs}
\usepackage{braket}

\newcommand{\be}{\begin{equation}}
\newcommand{\ee}{\end{equation}}
\newcommand{\ba}{\begin{eqnarray}}
\newcommand{\ea}{\end{eqnarray}}
\newcommand{\nn}{\nonumber}
\newcommand{\barr}{\begin{array}}
\newcommand{\earr}{\end{array}}

\newcommand{\bigoh}{\mathcal{O}}

\newcommand\lsim{\mathrel{\rlap{\lower4pt\hbox{\hskip1pt$\sim$}}
        \raise1pt\hbox{$<$}}}
\newcommand\gsim{\mathrel{\rlap{\lower4pt\hbox{\hskip1pt$\sim$}}
        \raise1pt\hbox{$>$}}}

\def\hE{{\widehat{\mathcal E}}}

\def\hH{{\widehat{\mathcal H}}}
\def\hT{{\widehat{\mathcal T}}}
\def\cL{{\mathcal L}}

\begin{document}

\title{New algorithms for radio pulsar search}

\author{Kendrick~M.~Smith}
\affiliation{Perimeter Institute for Theoretical Physics, Waterloo, ON N2L 2Y5, Canada}

\date{\today}

\begin{abstract}
The computational cost of searching for new pulsars is a limiting factor for
upcoming radio telescopes such as SKA.  We introduce four new algorithms: an optimal
constant-period search, a coherent tree search which permits optimal searching
with $\bigoh(1)$ cost per model,
a semicoherent search which combines information from coherent subsearches
while preserving as much phase information as possible, and a hierarchical search which
interpolates between the coherent and semicoherent limits.
Taken together, these algorithms improve the computational cost of pulsar search
by several orders of magnitude.  In this paper, we consider the simple case of a
constant-acceleration phase model, but our methods should generalize to
more complex search spaces.
\end{abstract}

\maketitle

\section{Introduction}

Pulsar science has been an exceptionally fertile area of astronomy over the last 50 years.
Some highlights include first observational evidence for gravitational waves~\cite{Taylor:1982zz},
first detection of extrasolar planets~\cite{Wolszczan:1992zg}, and exquisite tests of general
relativity from the double pulsar system~\cite{Lyne:2004cj}.

Although $\approx$2000 pulsars are known to date, the total observable population
is estimated to be larger by a factor $\sim$100~\cite{Yusifov:2004fr}.
Therefore, finding and timing pulsars is still in a relatively early stage.
Upcoming milestones include detection of nanohertz gravitational waves through pulsar 
timing~\cite{Detweiler:1979wn}, and discovery of a pulsar-black hole binary system, which would enable
new tests of strong gravity~\cite{Liu:2014uka}.

In practice, pulsar searching is very computationally expensive, and tradeoffs between
statistical optimality and computational cost are often necessary, particularly for upcoming instruments
such as SKA with large numbers of formed beams~\cite{Keane:2014vja}.
Because of this tradeoff, improving computational cost is more than a matter of convenience:
it means that searches can be more optimal, thus finding more pulsars.

We briefly summarize the current status of pulsar search algorithms.
Almost all pulsar searching is done using a variant of a power spectrum folding algorithm,
which can be described as follows.  
First suppose for simplicity that the pulse period is constant.
We square the Fourier transform of the data timestream $d(t)$ to obtain its power spectrum $P(\omega) = |\int d(t) e^{i\omega t} \, dt |^2$.
A constant-period pulsar shows up in the power spectrum as a sharp peak at the pulse frequency $\omega_p$, accompanied by
a series of decaying sharp peaks at higher harmonics $\omega = (n\omega_p)_{n\ge 2}$.
To sum these contributions, we ``fold'' the power spectrum, by computing $P_{\rm folded}(\omega) = \sum_{n=1}^{n_{\rm max}} W_n \, P(n\omega)$,
for some fixed weighting $W_n$, and take the maximum (over $\omega$) value of $P_{\rm folded}(\omega)$ as our search statistic.

If the pulsar frequency is not assumed constant, say for example a constant-acceleration model is assumed,
then one approach is to loop over trial accelerations, reparameterize the time coordinate to reduce
to the constant-period case, and compute the folded power spectrum of the reparameterized timestream.
Another approach is the Fourier-domain matched filter from~\cite{Ransom:2002zm}.

Power spectrum folding is not optimal, even if the weighting $W_n$ is matched to the power spectrum of the pulse profile.
Because pulses are narrow in the time domain, the phases of the Fourier modes $\tilde d(n\omega_p)$ are not independent, and 
power spectrum folding throws away this phase information.

Given infinite computational power, an optimal coherent search could be implemented by brute force as follows.
Assume a constant-acceleration model for simplicity.
We loop over trial acceleration $\alpha=\ddot\Phi$, trial frequency $\omega=\dot\Phi$, and trial phase $\Phi$.
For each triple $(\alpha,\omega,\Phi)$, we compute an intensity timestream $I_{\alpha\omega\Phi}(t)$,
and compute its overlap $\hE(\alpha,\omega,\Phi) = \int dt \, d(t) \, I_{\alpha\omega\Phi}(t)$ with the
data timestream $d(t)$.
This is manifestly optimal but computationally slow.
In this paper, we will show how to compute the transform $d(t) \rightarrow \hE(\alpha,\omega,\Phi)$
in a much faster way.
In the case where there are no trial accelerations (constant-period search), the transform simply factors
as a sequence of FFT's (\S\ref{sec:constp}).
The constant-acceleration search can be reduced to the constant-period case using a fast recursive
tree algorithm (\S\ref{sec:tree}).

The basic reason that pulsar searching is slow is that the size $S$ of the search
space, i.e.~the number of independent pulsar models in the search, is a rapidly 
growing function of the timestream length $T$.
For a constant-acceleration search, $S$ is proportional to one power of $T$ for $T \lsim T_a$,
and $S \propto T^3$ for $T \gsim T_a$, where $T_a$ is the threshold timestream length where
the constant-period approximation breaks down.
(For a pulsar with period $P$, period derivative $\dot P$, and duty cycle $D$, the threshold
is parametrically $T_a \sim P \, \dot P^{-1/2} D^{1/2}$.)
For large enough $T$, the constant-acceleration approximation will break down, and the
search space size will grow as $S \propto T^6$ if a single parameter $\ddot P$ suffices,
or even more rapidly if more parameters are needed.

In a strictly optimal coherent search, the overlap integral $\hE = \int dt \, d(t) I(t)$ is computed for
every model $I(t)$ in the search space.  A lower bound on the computational cost is $\bigoh(S)$, where
$S$ is the size of the search space, since $\bigoh(S)$ values of $\hE$ must be examined.  
Our recursive tree algorithm
in~\S\ref{sec:tree} saturates this bound, but for large timestreams, $S$ may still be too large
for the optimal search to be practical.  Therefore, we want to consider alternative search
algorithms which trade off optimality for speed.

The main idea of this paper is a proposal for such an algorithm, {\em semicoherent search} (\S\ref{sec:semicoherent}).
Our algorithm divides the data into chunks of size $T_c \ll T$, runs a coherent search in each chunk,
and combines these subsearches using a procedure which keeps as much phase information as possible.
The semicoherent search has an interpretation as the optimal search algorithm for a
phase model whose acceleration $\ddot\Phi$ is allowed to wander over a small range.
Even though the size of this search space is formally exponential in $T$, we will show that
its optimal search statistic satisfies recursion relations which permit evaluation in $\bigoh(T)$ time.
Intuitively, letting $\ddot\Phi$ wander ``fuzzes out'' the search space so that there is a finite
resolution to the phase information which must be retained, reducing computational cost.

Finally, we define a hierarchical search (\S\ref{sec:hierarchical}), which is simply a sequence of semicoherent 
searches with increasing coherence time $T_c$.  The most significant peaks from each search are passed
to the next search, as parameter ranges to be searched more optimally.
As $T_c$ increases, the semicoherent search becomes fully coherent.
Thus, if a pulsar is above a certain signal-to-noise threshold, then it will show up as a rare peak in each
level of the search, and the hierarchical search will converge to the complete phase model of the pulsar, as if a coherent
search had been run.
Of course, the key question is how this signal-to-noise threshold compares to a full coherent search,
or to other algorithms (such as power spectrum stacking).

The main result of this paper is in~\S\ref{sec:mc}, Fig.~\ref{fig:bottom_line}.
We study the efficiency of the hierarchical search in Monte Carlo simulations.
Remarkably, we find that the hierarchical search is nearly optimal for $T \lsim 64 T_c$.
In other words, given a timestream which is 64 times larger than the longest timescale
which can be searched coherently, we can still do a near-optimal search.
Since the hierarchical search has cost $\bigoh(T)$ and the coherent search
has cost $\bigoh(T^3)$, this should save a factor $64^2 = 4096$ in computing time.
For $T \gsim 64 T_c$, the hierarchical search is suboptimal, but the suboptimality grows
slowly, and the hierarchical search is closer to optimal than power spectrum stacking.

We note a few caveats.
First, our current code is a reference implementation and is not very well optimized.
Therefore, in this paper we will make rough estimates of computational cost, rather than giving hard timings.
We plan to improve this in follow-up work.

Second, we assume a fixed pulse profile throughout the analysis (a von Mises profile with duty cycle $D=0.1$).
In real data, we would need to add an outer loop over trial duty cycles.
Note that in a coherent search, it suffices to run the search at the smallest duty cycle, 
then obtain larger duty cycles by smoothing the output $\hE(\alpha,\omega,\Phi)$.
This procedure works because coherent search is a linear operation $d(t) \rightarrow \hE(\alpha,\omega,\Phi)$, 
but since the semicoherent search is nonlinear, we would need to rerun the search for each trial duty cycle.


Finally, throughout this paper, when we refer to a ``timestream'', we mean a dedispersed intensity
time series $I(t)$ at a fixed trial sky location and trial dispersion measure (DM).
The reader should keep in mind that the total computational cost of processing a survey is
larger by $(N_{\rm sky} N_{\rm DM})$, where $N_{\rm sky}$ is the number of sky pointings and $N_{\rm DM}$
is the number of trial DM's.  A full survey might have $N_{\rm sky} \sim 10^4$ pointings, and
$N_{\rm DM} \sim 1300$ or $N_{\rm DM} \sim 50000$ for a slow pulsar or millisecond pulsar search
respectively.  These $N_{\rm DM}$ values were derived assuming that the maximum DM of the search is 200 cm$^{-3}$ pc,
the spacing between trial DM's corresponds to a 1-sample delay across the full band, the survey has
100 MHz bandwidth at central frequency 400 MHz, and the sampling rate is $t_{\rm samp} = 2$ ms
for slow pulsars or $t_{\rm samp} = 50$ $\mu$s for millisecond pulsars.

\section{Preliminaries}

We model a pulsar by its phase model $\Phi(t)$ and pulse profile $\rho(\phi)$.
The phase model maps time $t$ to a dimensionless pulse phase $\Phi(t)$ such that the peak intensity
occurs when $\Phi(t)$ is a multiple of $2\pi$.
In this paper, the most general phase model we will consider is the constant-acceleration model, defined by
\be
\Phi(t) = \Phi_0 + \omega_0 t + \frac{1}{2} \alpha t^2  \label{eq:consta_model_intro}
\ee
with parameters $(t, \omega_0, \alpha)$.  The period $P$ and its derivative $\dot P$ are given by
$(P,\dot P) = (2\pi\omega^{-1}, -2\pi\alpha\omega^{-2})$.

The pulse profile $\rho(\phi)$ gives the pulse intensity as a function of pulse phase $\phi$.
Throughout this paper, we will assume the von Mises profile
\be
\rho(\phi) = e^{\kappa(\cos\phi-1)} = e^{-2\kappa \sin^2(\phi/2)}
\ee
where $\kappa$ is a concentration parameter.
Alternatively, we can parameterize the von Mises profile by its duty cycle $D$, which we define to be
the full width at half maximum (FWHM) of the pulse, divided by the pulse period.  The parameters $\kappa$
and $D$ are related by $\kappa = (\log 2) / (2 \sin^2(\pi D/2))$.

The intensity timestream $I(t)$ of the pulsar is simply the composite function $I(t) = \rho(\Phi(t))$.
We will assume that the timestream has been discretized with sample length $t_s$, so that the data
is a sequence $I_0, I_1, \cdots, I_{N-1}$ given by boxcar-averaging $I(t)$:
\be
I_k = \frac{1}{t_s} \int_{kt_s}^{(k+1)t_s} dt \, I(t)  \label{eq:boxcar}
\ee
We will assume that the noise is Gaussian and uncorrelated,
i.e. the noise covariance is $\langle I_k  I_l \rangle = \eta^2 t_s^{-1} \delta_{kl}$.
Here, $\eta$ is a noise parameter with units intensity-(time)$^{1/2}$.

We write $\bar I_k$ for the discretized signal normalized to signal-to-noise 1,
i.e.~obeying normalization condition
\be
\eta^{-2} t_s \sum_{k=0}^{N-1} (\bar I_k)^2 = 1  \label{eq:Ibar_normalization}
\ee
Note that the normalization of $\bar I_k$ implicitly depends on the number of timestream samples $N$.

Given data realization $d_k$ and fixed pulsar model $I_k$, the optimal statistic for detecting the
pulsar is the coherent detection statistic $\hE$ defined by:
\be
\hE = \eta^{-2} t_s \sum_{k=0}^{N-1} d_k \bar I_k  \label{eq:Edef}
\ee
By ``coherent'', we mean that all phase information is used.
We have normalized $\hE$ so that its numerical value is the detection significance in sigmas.
Equivalently, $\hE$ is equal to $(\Delta\chi^2)$, the improvement in $\chi^2$ after subtracting
a best-fit multiple of the pulsar model $\bar I_k$.
This normalization is convenient, but note that if two timestreams of lengths $N_1, N_2$ are combined, 
the rule for combining $\hE$-values is:
\be
\hE = \frac{N_1^{1/2}}{(N_1+N_2)^{1/2}} \hE_1 + \frac{N_2^{1/2}}{(N_1+N_2)^{1/2}} \hE_2  
\ee

\section{Fast constant-period search}
\label{sec:constp}

In this section we will consider the simplest possible search space: a pulsar with constant frequency,
parameterized by a frequency $\omega$ and phase $\Phi$.
A brute-force optimal search algorithm for this search space would be to loop over a grid of trial parameters $(\omega,\Phi)$,
and compute the coherent statistic $\hE$ defined in Eq.~(\ref{eq:Edef}), which becomes a function $\hE(\omega,\Phi)$.
In this section we will show that the brute-force search can be computed in a mathematically equivalent but faster way.
If we view the coherent search as a transform $d(t) \rightarrow \hE(\omega,\Phi)$, then the transform can be factored
as a sequence of FFT's.

The phase model for this search is:
\be
\Phi(t) = \Phi_c + \omega_c \left( t - \frac{T}{2} \right)  \hspace{1.5cm} (0 \le t \le T) \label{eq:constp_phase_model}
\ee
where $T = Nt_s$ is the total timestream length, $\omega_c$ is the angular pulse frequency,
and $\Phi_c$ is the pulse phase at the timestream center $t=T/2$.
Note that we have taken the model parameters to be frequency and central phase $\Phi_c$,
whereas previously in Eq.~(\ref{eq:consta_model_intro}) we used the initial phase 
$\Phi_0 = \Phi_c - \omega T/2$ instead of $\Phi_c$.
This change of variables will be convenient for reasons to be explained shortly.

We Fourier transform the pulse profile $\rho(\phi)$:
\be
\rho(\phi) = \sum_n \rho_n e^{in\phi}
\ee
where the sum runs over positive and negative $n$.
We then write the signal timestream $I(t)$ as
\ba
I(t) &=& \rho(\Phi(t)) \nn \\
  &=& \rho\!\left( \Phi_c + \omega_c\left(t - \frac{T}{2} \right) \right) \nn \\
  &=& \sum_n \rho_n e^{in\Phi_c} e^{in \omega_c (t-T/2)}  \label{eq:constp_nondisc}
\ea
The discretized timestream $I_k$ is obtained from the continuous timestream $I(t)$ by boxcar averaging.
Equivalently, we can convolve $I(t)$ with a length-$t_s$ boxcar, then evaluate at the sample center $t = (k+1/2) t_s$.
The convolution can be implemented by multiplying each Fourier mode of $I(t)$ by $j_0(\omega t_s/2)$, the Fourier
transform of a length-$t_s$ boxcar.  Therefore, starting from Eq.~(\ref{eq:constp_nondisc}) the discretized timestream can be written:
\be
I_k = \sum_n \rho_n \, j_0\!\left( \frac{n\omega_c t_s}{2} \right) e^{in\Phi_c} e^{in \omega_c (kt_s+t_s/2-T/2)}  \label{eq:constp_signal}
\ee
where $j_0(x)=(\sin x)/x$.

The normalized timestream $\bar I_k$ defined in the previous section is given by 
$\bar I_k = A^{-1/2} I_k$, where the normalization $A$ is given by:
\be
A = \eta^{-2} t_s \sum_k (I_k)^2
\ee
We next derive an approximate formula for $A$, in the limit where the timestream length $T$
is large compared to the pulse period.
We approximate the sum, which has the schematic form $\eta^{-2} t_s \sum_k I(kt_s)^2$, by the integral $\eta^{-2} \int dt \, I(t)^2$,
and plug in Eq.~(\ref{eq:constp_signal}) to obtain:
\be
A \approx \eta^{-2} \int dt \, \left[ \sum_n \rho_n j_0\!\left( \frac{n\omega_c t_s}{2} \right) e^{in\Phi_c} e^{in \omega_c (t+t_s/2-T/2)} \right]^2
\ee
We now expand the square, making the approximation $\int dt \, e^{im\omega_c t} e^{in\omega_c t} \approx T \delta_{m,-n}$.
This gives:
\be
A(\omega_c) \approx \eta^{-2} \, T \sum_n |\rho_n|^2 \, j_0\!\left( \frac{n\omega_c t_s}{2} \right)^2  \label{eq:Aomega}
\ee
where we have written $A(\omega_c)$ on the LHS to emphasize that it depends on $\omega_c$ but not $\Phi_c$.
We can use this formula to get the correct normalization for $\bar I_k$, starting from an arbitrary
normalization for the pulse profile $\rho_n$.

Now we obtain a formula for the search statistic $\hE$, by plugging 
Eq.~(\ref{eq:constp_signal}) into the definition~(\ref{eq:Edef}) of $\hE$:
\be
\hE(\omega_c,\Phi_c) = A(\omega_c)^{-1/2} \eta^{-2} t_s \sum_k d_k \sum_n \rho_n j_0\!\left( \frac{n\omega_c t_s}{2} \right) e^{in\Phi_c} e^{in \omega_c (kt_s+t_s/2-T/2)}
\ee
To simplify this, we define the Fourier transform of the data realization $d_k$ by:
\be
\tilde d(\omega) = t_s \, \sum_k d_k e^{i\omega(kt_s+t_s/2-T/2)}   \label{eq:domega_def}
\ee
Note that this definition contains an extra phase $e^{i\omega(t_s/2-T/2)}$ relative to the usual Fourier transform.
Then $\hE(\omega_c,\Phi_c)$ can be written
\be
\hE(\omega_c,\Phi_c) = \frac{1}{A(\omega_c)^{1/2} \eta^2} \sum_n \rho_n j_0\!\left( \frac{n\omega_c t_s}{2} \right) \tilde d(n\omega_c) e^{in\Phi_c}  \label{eq:constp_fast}
\ee
This formula can be used to give a fast algorithm for evaluating $\hE(\omega_c,\Phi_c)$.
First, we precompute $\tilde d(\omega)$ on a grid of $\omega$-values, 
by zero-padding the timestream and taking an FFT.
The purpose of the zero-padding is to make the $\omega$-sampling narrow enough that $\tilde d(\omega)$ can be
evaluated at an arbitrary frequency by interpolation.
We find that zero-padding by a factor two is sufficient (Appendix~\ref{app:spacings}).
Similarly, we precompute $\rho_n$, and populate an interpolation table with precomputed values of $A(\omega)$
using Eq.~(\ref{eq:Aomega}).

Now to evaluate $\hE(\omega_c,\Phi_c)$ on a grid of trial $(\omega_c,\Phi_c)$ values, we 
evaluate Eq.~(\ref{eq:constp_fast}) by interpolating the factors which appear on the RHS.
To do the $n$-sum efficiently, we use an FFT from the variable $n$ to variable $\Phi_c$.

This concludes our fast FFT-based algorithm for constant-frequency search, but there
are a few details which deserve elaboration.
We have chosen to use the central phase $\Phi_c$ as a model parameter in Eq.~(\ref{eq:constp_phase_model}),
rather than the initial phase $\Phi_0$.  This resulted in an phase shift $e^{-i\omega(T/2)}$ in the 
definition~(\ref{eq:domega_def}) of $\tilde d(\omega)$.  Empirically, we find that this choice results
in a more well-behaved $\tilde d(\omega)$ interpolation.  This can also be understood formally by
noticing that the phase $e^{i\omega(kt_s+t_s/2-T/2)}$ appearing in Eq.~(\ref{eq:domega_def}) has fewer
wraparounds (as $\omega$ is varied) with the factor $e^{-i\omega T/2}$ than without.

There are a hidden parameters in our algorithm: the amount of zero-padding used
to compute the timestream FFT, and the spacings of the trial parameters $\Phi_c$ and $\omega_c$.
Strictly speaking, our algorithm is only optimal in the limit of large zero-padding and
small trial spacings.
In practice we choose parameter values which are a compromise between optimality and computational cost.
In Appendix~\ref{app:spacings} we give a formal procedure for making these choices and
recommend some default values.

In real data analysis, the timestream $d_k$ must be ``detrended'' or high-pass filtered before
being searched for pulsars.  It would be equivalent to leave the data $d_k$ unfiltered, but apply
high-pass filtering to the pulsar signal $\bar I_k$ before computing the search statistic $\hE$.
Therefore, we can account for the effect of detrending by simply subtracting the 
mean from the pulse profile $\rho(\phi)$, or equivalently setting the Fourier mode $\rho_0$ to zero.
All results in this paper include this detrending correction, but it turns out to make little difference.

The computational cost of our fast algorithm can be estimated as follows.
The initial FFT has cost $\bigoh(N \log N)$, where $N$ is the number of time samples,
and the $\hE$ calculation has cost $\bigoh(N_\omega N_\Phi \log N_\Phi)$, where
$N_\omega, N_\Phi$ are the number of trial $\omega$ and $\Phi$ values needed.
It is not hard to see that $N_\Phi = \bigoh(D^{-1})$ and $N_\omega = \bigoh(\omega_{\rm max} T D^{-1})$,
where $D$ is the duty cycle and $\omega_{\rm max}$ is the maximum value of $\omega$ which is searched.
We will assume that the data has been sampled so that the sample length $t_s$ is comparable to the
minimum pulse width in the search, or equivalently $N = \bigoh(\omega_{\rm max} T D^{-1})$.
Putting all of this together, we can write the total cost as $\bigoh(N \log N + N D^{-1} \log D^{-1})$.

This cost can be compared to the power spectrum folding algorithm described in the introduction,
which has cost $\bigoh(N \log N)$.  The computational costs will typically be comparable, but our
coherent algorithm will be significantly slower in the limit of low duty cycle.
On the other hand, this is also the limit where power spectrum folding becomes significantly 
suboptimal, so one could argue that it is always a good idea to use the fast coherent search instead of 
power spectrum folding.

\section{Tree algorithm for constant-acceleration search}
\label{sec:tree}

We now consider a more complex case: a constant-acceleration search.
In this case, the output of the coherent search will be a function of three variables $\hE(\alpha,\omega,\phi)$,
where $\alpha=\ddot\Phi$ is an acceleration parameter.  We will give a fast algorithm for evaluating $\hE$ on
a grid of trial parameters.

The algorithm is recursive and based on the following idea.
We divide the time interval $[0,T]$ into two subintervals $[0,T/2]$ and $[T/2,T]$.
At any point $(\alpha,\omega,\phi)$, the statistic $\hE$ will be a sum of contributions $\hE = \hE_1 + \hE_2$.
The number of grid points needed to fully represent $\hE_1$ and $\hE_2$ will smaller by a factor $(1/8)$,
since the number of trial accelerations scales as $T^2$ and the number of trial frequencies scales as $T$.
Therefore we can compute $\hE_1$ and $\hE_2$ on coarser grids, and interpolate to a finer grid when
we sum them to obtain $\hE$.
The values of $\hE_1$ and $\hE_2$ are computed recursively using the same method, further subdividing the time range.
After enough subdivisions, the timestream will be short enough that no trial accelerations are needed,
and the fast constant-period search from the previous section can be used to compute $\hE$,
ending the recursion.

We parameterize the phase model as:
\be
\Phi(t) = \bar \Phi + \omega_c \left( t - \frac{T}{2} \right) + \frac{1}{2} \alpha \left[ \left( t - \frac{T}{2} \right)^2 - \frac{T^2}{12} \right]  \label{eq:consta_leg}
\ee
over the range $0 \le t \le T$.  Note that we have changed variables from the parameterization
$(\alpha,\omega_0,\Phi_0)$ given previously in Eq.~(\ref{eq:consta_model_intro}) to the parameters
$(\alpha,\omega_c,\bar\Phi)$.  The parameter $\omega_c$ is the derivative $d\Phi/dt$ evaluated
at the central time $t=T/2$, and $\bar\Phi$ is the mean value of $\Phi$ over the range $0\le t\le T$,
as suggested by the notation.
An analogous change of variables was made in the last section (Eq.~(\ref{eq:constp_phase_model})), 
in order to make an interpolation better behaved.
The motivation here is similar and will be described shortly.

Let $\hE_1, \hE_2$ denote the search statistic $\hE$ restricted to the subinterval $[0,T/2]$ or $[T/2,T]$.
When we write $\hE_1(\alpha,\omega_c,\bar\Phi)$, the arguments $(\omega_c,\bar\Phi)$ are always
defined relative to the subinterval $[0,T/2]$, not the larger interval $[0,T]$ (and likewise
for $\hE_2$).
In this notation, a short calculation gives the recursion relating $\hE$ to $\hE_1,\hE_2$:
\be
\hE(\alpha,\omega_c,\bar\Phi) = 
   \frac{1}{\sqrt{2}} \hE_1\!\left(\alpha,\omega_c-\frac{\alpha T}{4}, \bar\Phi-\frac{\omega_c T}{4}\right)
+  \frac{1}{\sqrt{2}} \hE_2\!\left(\alpha,\omega_c+\frac{\alpha T}{4}, \bar\Phi+\frac{\omega_c T}{4}\right)  \label{eq:tree}
\ee
Now suppose that the search statistics $\hE_1,\hE_2$ have been precomputed on a complete grid of trial 
$(\alpha,\omega_c,\bar\Phi)$ values.
We use the recursion relation~(\ref{eq:tree}) to evaluate $\hE$ on a complete grid,
using interpolation to obtain values of $\hE_1,\hE_2$ off-grid.
The precomputation of $\hE_1$ and $\hE_2$ is done by applying the same idea recursively, initializing $\hE_1$
from tables $\hE_{11}$ and $\hE_{12}$ which correspond to intervals $[0,T/4]$ and $[T/4,T/2]$, and so on.
After enough subdivisions, the subinterval size $(T/2^N)$ is small enough that the acceleration $\alpha$
is negligible, and we can approximate $\hE(\alpha,\omega_c,\bar\Phi) \approx \hE(0,\omega_c,\bar\Phi)$.
At this point, the grid of $\hE$ values can be computed using the constant-period algorithm from the previous section,
which ends the recursion.

The computational cost of the search can be computed as follows.
If we write the total number of $\hE$ evaluations needed at the top level of the search as $N_\alpha N_\omega N_\Phi$,
then the total number of $\hE_1$ and $\hE_2$ evaluations needed at the second level is 
$2 (N_\alpha/4) (N_\omega/2) N_\Phi = N_\alpha N_\omega N_\Phi/4$.
Similarly, the total number of evaluations needed at the third level of the search is $N_\alpha N_\omega N_\Phi/16$, and so on.
Summing over all levels of the search, the total computational cost is $\bigoh(N_\alpha N_\omega N_\Phi + N \log N + N D^{-1} \log D^{-1})$,
where the first term is the sum of a convergent geometric series, and the second and third terms are the total cost of all the
constant-period searches at the bottom level.  Provided that $N_\alpha \gsim \log N$ and $N_\alpha \gsim D^{-1} \log D^{-1}$, which will be the case
for all large constant-acceleration searches, the first term dominates and we can write the total cost as
$\bigoh(N_\alpha N_\omega N_\Phi)$.
Remarkably, we see that the computational cost of the tree search is $\bigoh(1)$ per model in the search space!

The tree algorithm contains hidden parameters: the grid spacings in the parameters
$(\alpha,\omega_c,\bar\Phi)$ used to construct interpolation tables at each level of the search,
and the threshold for switching to the bottom-level constant-period search.
On a related note, the reader may wonder whether there is any suboptimality introduced cumulatively
by the chain of interpolations in the tree search.
In Appendix~\ref{app:spacings}, we show that it is straightforward to choose parameters so that
the search is as close to optimal as desired.
We give our default parameter choices, and show that they produce a search which is $> 90\%$ optimal.

A very interesting aspect of the tree algorithm is that it should generalize straightforwardly
to search spaces more complex than the constant-acceleration search, for example a polynomial
search of degree 3 or 4.  As the search recurses, trial parameter spacings can be decreased, and
the polynomial degree can also be decreased.
We defer an exploration of more complex search spaces to future work.

The parameterization $(\alpha,\omega_c,\bar\Phi)$ was chosen in Eq.~(\ref{eq:consta_leg})
because we find empirically that the interpolation $\hE(\alpha,\omega_c,\bar\Phi)$ is better behaved,
than a simpler parameterization such as $(\alpha,\omega_0,\Phi_0)$.
In fact, the parameterization in Eq.~(\ref{eq:consta_leg}) is just the Legendre polynomial
expansion truncated at degree 2 and rescaled to the interval $[0,T]$.
Since Legendre polynomials are orthogonal, varying their coefficients produces
an uncorrelated effect on $\hE$, which leads to a more efficient interpolation.
This way of thinking about the parameterization makes the generalization to higher-degree
polynomial models transparent.

So far, we have discussed searches which are strictly optimal, in the sense that the optimal statistic $\hE$
is evaluated for every model in the search space.
The tree algorithm from this section has computational cost $\bigoh(S)$, where $S = N_\alpha N_\omega N_\Phi$
is size of the search space, i.e.~the number of independent phase models.

It may seem plausible that this is a lower bound on the computational cost of any optimal search, by
the following argument.
Even if $\hE$ had been precomputed for every model and stored in memory, we still need to inspect $S$
values of $\hE$ in order to perform the search, and the cost of simply reading these values from memory
is $\bigoh(S)$.
Surprisingly, this simple argument is incorrect!  As we will see in~\S\ref{sec:semicoherent}, there are examples
of ``fuzzy'' search spaces which can be optimally searched with computational cost $\bigoh(\log S)$, where $S$ is
the size of the space.
Although a constant-acceleration search is not an example of a fuzzy search space, there is a sense in
which it can be approximated by one.  This will lead us to the notion of a semicoherent search, 
the main idea of this paper.

\section{The $\hH$-statistic}

In this section, we study the following question.  Consider a discrete search space which consists
of $S$ distinct pulsar models $I_k^{(1)}, \cdots, I_k^{(S)}$, and suppose we have evaluated $\hE$ for each
model, obtaining $S$ values $\hE_1, \cdots, \hE_S$.  If we want to compress these $S$ numbers into a single
number which represents the result of the search, what should we do?  Should we use the max-statistic
$\max(\hE_1,\cdots,\hE_S)$, the $\chi^2$-like statistic $\sum_{s=1}^S (\hE_s)^2$, or something else?

Let us interpret this question in the language of frequentist hypothesis testing.
We would like to compare two hypotheses, a null hypothesis that the data is pure noise, and an alternative
hypothesis that the data is noise plus a pulsar signal.
To specify the alternative hypothesis precisely, let us suppose that $r \bar I_k^{(s)}$ is added to the
noise, where $r$ is a fiducial signal-to-noise in sigmas, and $s=1,\cdots,S$ is a random pulsar model.

By the Neyman-Pearson lemma, the optimal statistic for distinguishing these hypotheses is
the likelihood ratio $(\cL_1/\cL_0)$, where $\cL_0$ and $\cL_1$ are the likelihoods of the data given the null
and alternative hypotheses.  To write these likelihoods compactly, we introduce a dot product notation.
Given two discretized timestreams $x_k, x'_k$, we define their dot product by:
\be
x \cdot x' = \eta^{-2} t_s \sum_k x_k x'_k  \label{eq:dot_def}
\ee
In this notation, the estimator $\hE$ is defined by $\hE = d \cdot \bar I$, and the normalization of $\bar I$
defined in Eq.~(\ref{eq:Ibar_normalization}) is $\bar I \cdot \bar I = 1$.
Neglecting overall constants, the likelihoods $\cL_0$ and $\cL_1$ are given by
\be
\cL_0 \propto e^{-(d\cdot d)/2}
  \hspace{2cm}
\cL_1 \propto \frac{1}{S} \sum_{s=1}^S e^{-(d-r\bar I^{(s)}) \cdot (d-r\bar I^{(s)})/2}
\ee
After a little algebra, the likelihood ratio can be written as $(\cL_1/\cL_0) \propto e^{r\hH}$,
where we define the $\hH$-statistic by:
\be
\hH = \frac{1}{r} \log \frac{1}{S} \sum_{s=1}^S e^{r\hE_s}  \label{eq:Fhat_def}
\ee
We can think of $\hH$ as a version of $\hE$ which has been coarse-grained over a model space
of size $S$.  Note that the definition of $\hH$ depends both on the model space and the fiducial
signal-to-noise $r$.  The statistic $\hH$ is only strictly optimal if $r$ is equal to the true
signal-to-noise of the pulsar.  In practice, we set $r$ equal to the detection threshold of the
search.  It makes sense to optimize the search for pulsars near threshold, since pulsars with
signal-to-noise significantly above threshold will still be detected if the statistic is slightly
suboptimal, and pulsars significantly below threshold will never be detected.

Note that $\hH$ is a likelihood ratio test in disguise, since it is just a reparameterization
of $(\cL_1/\cL_0)$.  This reparameterization is convenient because the value of $\hH$
on a timestream containing a bright pulsar (i.e.~significantly above threshold) is simply its
coherent signal-to-noise in sigmas.  We also note that $\hH = \hE$ if the search space size $S=1$.

We have now answered the question from the beginning of this section: the statistic $\hH$
defined in Eq.~(\ref{eq:Fhat_def}) compresses the values $\hE_1,\cdots,\hE_S$ into a single
optimal statistic.
As an aside, we note that $\hH$ interpolates between the max-statistic 
$\max(\hE_1,\cdots,\hE_S)$ and the $\chi^2$-like statistic $\sum_{s=1}^S (\hE_s)^2$
in the limits $r\rightarrow \infty$ and $r\rightarrow 0$ respectively.
The first statement is easy to show, and the second statement follows by Taylor expanding
\be
\hH = \frac{1}{S} \sum_s \hE_s - \frac{r}{2S^2} \Big( \sum_s \hE_s \Big)^2 + \frac{r}{2 S} \sum_s (\hE_s)^2 + \bigoh(r^2)  \label{eq:fhat_taylor}
\ee
We write the quantity $\sum_s \hE_s$ as $d \cdot (\sum_s \bar I_s)$, and note that $\sum_s \bar I_s$ vanishes
if the pulsar profile is detrended and the search space is invariant under the phase shift symmetry
$\Phi \rightarrow \Phi + \Delta\Phi$.  Therefore the first two terms in the Taylor series~(\ref{eq:fhat_taylor}) vanish,
and the leading term as $r \rightarrow 0$ is the third term, which is proportional to the $\chi^2$-like 
statistic $\sum_{s=1}^S (\hE_s)^2$.

\section{Semicoherent search}
\label{sec:semicoherent}

In this section, we introduce the notion of a semicoherent search.
Suppose we are interested in a constant-acceleration search over acceleration range $[-\alpha_0, \alpha_0]$.
We divide the timestream into $N_c \gg 1$ chunks, where the chunk duration $T_c = T/N_c$ is assumed
comparable to the acceleration timescale $T_a = (D/\alpha_0)^{1/2}$.
Under these assumptions, a constant-period search would suffice if we restrict the search to a single length-$T_c$
chunk, but we would need to use multiple trial $\alpha$-values to search larger chunk sizes.

Let us imagine that it is computationally feasible to do a constant-period search in each length-$T_c$ chunk,
but we do not have enough computing power to analyze larger chunk sizes.
How can we stitch together the per-chunk search statistics $\hE_i(\omega_c,\Phi_c)$, where $i=1,\cdots,N_c$,
in order to form the best possible search statistic?
Of course, this will be a suboptimal statistic, since we are assuming that 
the fully coherent search is not affordable, but we are asking for the {\em least suboptimal} way of assembling
coherent searches computed on the timescale $T_c \ll T$.

Informally, the idea of the semicoherent search is to compute the $\hH$-statistic for a ``fuzzy'' search space
consisting of all phase models whose acceleration $\alpha = \ddot\Phi$ is allowed to wander over the 
range $[-\alpha_0, \alpha_0]$.
In each length-$T_c$ chunk, the constant-period approximation will be valid.
The $\hH$-statistic for the fuzzy search space will combine the per-chunk $\hE$ values in a way which
keeps as much phase consistency between the chunks as possible.

Formally, we introduce the following search space.
We consider phase models which have the quadratic form $\Phi(t) = (1/2) \alpha t^2 + \omega t + \Phi_0$
in each chunk, but we allow the acceleration $\alpha = \ddot\Phi$ to change discontinuously at chunk boundaries.
Specifically, we assume that $\alpha = \pm \alpha_0$, where the sign is independent in each chunk.
We require the phase $\Phi$ and its derivative $\dot\Phi$ to be continuous across chunk boundaries.
Thus we can parameterize the search space by initial phase $\Phi_0$, an initial frequency $\omega_0$,
and a sequence of signs $s_i = \pm 1$, where $1 \le i \le N_c$.
This data suffices to determine the phase model $\Phi(t)$ for all times $0 \le t \le T$.

The size of this search space grows exponentially with $T$, so it would be natural to assume that the computational
cost of searching it is also exponential.  However, we will now show that there is a recursion relation which
allows the $\hH$-statistic for this search space to be computed in linear time.

Given a pulsar in the search space, the coherent search statistic $\hE$ will be a sum over contributions
from each chunk.  Schematically, $\hE = (\hE_1 + \cdots + \hE_{N_c}) / \sqrt{N_c}$.
Now suppose we define an $\hH$-statistic by summing over all $2^{N_c}$ phase models with
initial phase and frequency $(\Phi_0, \omega_0)$.  We can write $\hH$ as an iterated sum:
\be
\hH(\omega_0,\Phi_0) = \frac{1}{r} \log \sum_{s_1} \cdots \sum_{s_{N_c}} \exp\left( \frac{r}{\sqrt{N_c}} \hE_1 + \cdots \frac{r}{\sqrt{N_c}} \hE_{N_c} \right)  \label{eq:sc_H1}
\ee
where it is understood that each $\hE_i$ on the RHS is evaluated on an appropriate set
of model parameters, which depend on $(\omega_0,\Phi_0)$ and the signs $s_j=\pm 1$ with $j \le i$.

Note that the definition in Eq.~(\ref{eq:sc_H1}) partially coarse-grains the search space: we have
coarse-grained over the signs $s_i$, but not the phase $\Phi_0$ or frequency $\omega_0$.

We now write a recursion relation which allows $\hH$ to be computed efficiently.
Given a chunk index $0 \le j < N_c$, and a frequency and phase $(\omega_j,\Phi_j)$
defined at the intermediate time $t=jT_c$, we define the partially summed statistic 
$\hH_j(\omega_j,\Phi_j)$ by:
\be
\hH_j(\omega_j,\Phi_j) = \frac{1}{r} \log \sum_{s_{j+1}} \cdots \sum_{s_{N_c}} \exp\left( \frac{r}{\sqrt{N_c}} \hE_{j+1} + \cdots \frac{r}{\sqrt{N_c}} \hE_{N_c} \right)  \label{eq:sc_H2}
\ee
We have coarse-grained over the $2^{N_c-j}$ choices of sign which connect the intermediate time $jT_c$ 
and the final time $T = N_cT_c$.

There is a recursion relation relating $\hH_j$ and $\hH_{j+1}$.  To see this, note that if we fix the index $s_{j+1}$ in
the outermost sum in Eq.~(\ref{eq:sc_H2}), then the iterated sum which remains is the same sum which appears in the definition
of $\hH_{j+1}$.  More precisely, a short calculation gives the recursion:
\be
\hH_j(\omega_j,\Phi_j) = \frac{1}{r} \log \sum_{s=\pm 1} \exp \left[ 
    \frac{r}{\sqrt{N_c}} \hE_j\!\left(\omega_j + s \frac{\alpha_0 T_c}{2}, \Phi_j + \frac{\omega_j T_c}{2} + s \frac{\alpha_0 T_c^2}{6} \right)
  + r \hH_{j+1}\!\left( \omega_j + s \alpha_0 T_c, \Phi_j + s \frac{\alpha_0 T_c^2}{2} \right)   \label{eq:sc_Hrec}
 \right]
\ee
This recursion can be used to compute $\hH_j$ in the order $j=N_c,\cdots,0$.
The initial condition for the recursion is $\hH_{N_c}(\omega,\Phi)=0$.
The output of the semicoherent search is a 2D array of values $\hH_0(\omega_0,\Phi_0)$
with shape $(N_\omega, N_\Phi)$, where $N_\omega$ is the number of trial frequencies
associated with timestream length $T_c$ (not length $T=N_cT_c$).
The computational cost of iterating the recursion is $\bigoh(N_c N_\omega N_\Phi)$.
For a fixed chunk size $T_c$, the cost grows as one power of $T$,
in contrast to a coherent search, whose cost would grow as $T^3$.

\section{Hierarchical search}
\label{sec:hierarchical}

First, we introduce a small generalization of the semicoherent search.
In the previous section, we defined a single acceleration bin 
$[-\alpha_0, \alpha_0]$, and allowed $\ddot\Phi$ to wander over this range.  
As a generalization, suppose we cover the acceleration range of the search with $N_\alpha$
acceleration bins, with bin width $(\Delta\alpha)$.  We define a multi-bin semicoherent
search, by modifying the single-bin construction above as follows:
\begin{itemize}
\item We choose the chunk size $T_c$ comparable to the timescale $(D/\Delta\alpha)^{1/2}$ where the trial acceleration spacing
  of a coherent search is $(\Delta\alpha)$.  This will be a longer timescale than in the single-bin case,
  where $T_c \sim T_a = (D/\alpha_0)^{1/2}$ is the timescale where no trial accelerations are needed at all.
\item In each length-$T_c$ chunk, we run a coherent search, which will have $N_\alpha$ trial accelerations,
  using the recursive tree algorithm from~\S\ref{sec:tree}.  (In the single-bin case, we chose $T_c$ so
  that no trial accelerations were necessary, and used the fast constant-frequency search algorithm from~\S\ref{sec:constp}.)
\item For each acceleration bin $a$, initial frequency $\omega_0$, and initial phase $\Phi_0$,
  we coarse-grain over all phase models whose acceleration $\ddot\Phi$ is allowed to wander over
  the bin $a$.  The resulting $\hH$-statistic $\hH(a,\omega_0,\Phi_0)$ can be computed using the
  recursion relation~(\ref{eq:sc_Hrec}).
\item The output of the semicoherent search is a 3D array of values $\hH(a,\omega_0,\Phi_0)$
  with shape $(N_\alpha, N_\omega, N_\phi)$.  Here, $N_\alpha$ is the number of acceleration bins
  which were specified as an input parameter of the search, and $N_\omega$ is the number of trial
  frequencies associated with timestream length $T_c$.  The computational cost is 
  $\bigoh(N_\alpha N_\omega N_\Phi)$.
\end{itemize}
We can think of the semicoherent search as being parameterized by an $\alpha$-bin width $(\Delta\alpha)$,
or by a coherence length $T_c$.
The two will be related by $T_c \propto (D/\Delta\alpha)^{1/2}$, but the constant of proportionality
is a parameter that can be optimized.
In practice, we usually determine $T_c$ from considerations of computational cost: we simply set it
to the largest chunk size where we can afford to do a coherent search.
The semicoherent search can be interpreted as a procedure for stitching together the results of coherent
searches on the timescale $T_c \ll T$, to obtain a single search statistic $\hH$ which preserves as much
phase information as possible.

One more definition.  A ``hierarchical search'' is just a sequence of semicoherent searches with
increasing values of $T_c$.  (In our implementation, we take each $T_c$ to be 4 times larger than
the previous level of the hierarchy.)  At each level of the hierarchy, the rare peaks in the output
array $\hH(a,\omega_0,\Phi_0)$ are used to define input search ranges for the search at the next level
of hierarchy.  Because only a small fraction of the search space is passed on, the total computational 
cost in our implementation is dominated by the first level of the hierarchy.  

The last level of the hierarchy is a coherent search, which only runs on a very small fraction of the 
full $(\alpha,\omega,\Phi)$ parameter space.  If the timestream contains a bright pulsar, then it will produce 
a rare peak at every level of the hierarchy, and the hierarchical search will return a fit for the full phase
model, as if a coherent search had been run.  Of course, the key question is, what is the signal-to-noise
threshold for the hierarchical search to work?  In the next section, we study this question using 
Monte Carlo simulations.

\section{Monte Carlo simulations}
\label{sec:mc}

In this section, we run Monte Carlo simulations using the following fiducial survey parameters.
We assume time sampling $t_{\rm samp}=50$ $\mu$sec as appropriate for millisecond pulsars, 
and total timestream size $2^{28}$ samples, corresponding to integration time $T = 3.7$ hours.
Our search assumes a von Mises profile with fixed duty cycle $D = 0.1$.

We search the frequency range $(\omega_{\rm min}, \omega_{\rm max}) = (200, 4000)$ Hz,
corresponding to pulsar period $P$ between 1.5 and 31 milliseconds.
We search acceleration range $(\alpha_{\rm min}, \alpha_{\rm max}) = (-10^{-4}, 10^{-4})$ sec$^{-2}$.
For relativistic binary pulsars with $v/c=10^{-3}$, this acceleration range covers
orbital periods $P_{\rm orb}$ greater than 3.5 hours if $P=31$ msec,
or $P_{\rm orb}$ greater than 70 hours if $P=1.5$ msec.\footnote{Our fiducial survey
parameters are not completely consistent, since a binary system with $|\ddot\Phi| \sim 10^{-4}$
would not be fit by a constant-acceleration model over observation time 3.7 hours.
In future work, we plan to extend our methods to searches more complex than a
constant-acceleration search.  In this paper, the fiducial survey is just intended
as a way of comparing our method to its alternatives in a computation-limited
regime.}


With these parameters, we estimate that a fully coherent search would have search space 
size $S = N_\alpha N_\omega N_\phi \approx 2 \times 10^{13}$.
This represents the computational cost of a single sky pointing and trial DM.
In a full millisecond pulsar survey, the cost would be larger by a factor 
$N_{\rm sky} N_{\rm DM} \sim 10^9$ as described in the introduction.
Therefore, a full coherent search is not computationally feasible for a full
survey with these parameters.

\begin{figure}
\centerline{\includegraphics[width=14cm]{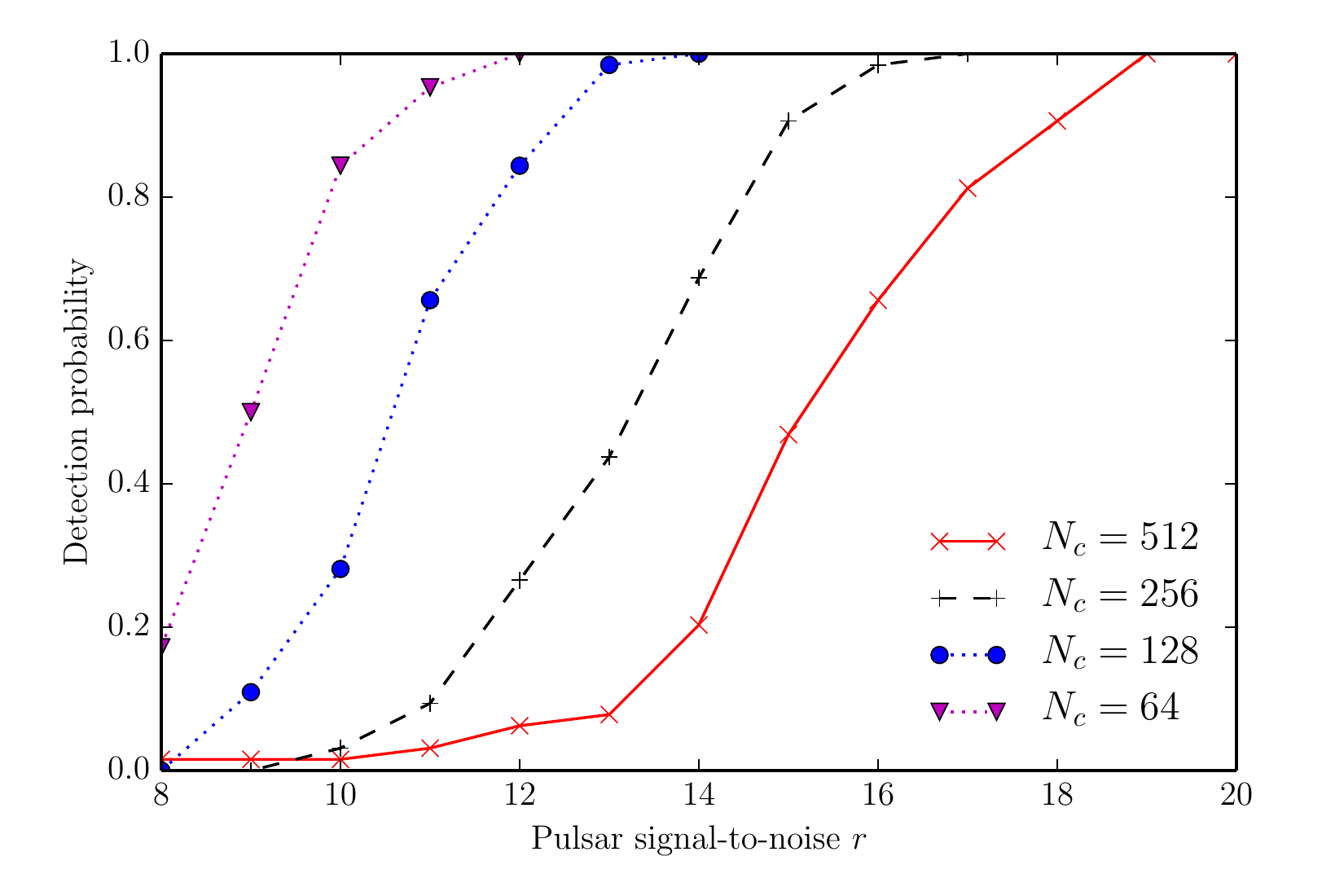}}
\caption{Monte Carlo detection probability of hierarchical search, as a function of 
 the number of coherent chunks $N_c$ and the total signal-to-noise $r$ of the simulated pulsar.
 In the text, we interpret these detection probabilities and show that the hierarchical search
 is close to optimal at $N_c=64$.  For larger $N_c$ it is less suboptimal than incoherent power 
 spectrum stacking, an alternative chunk-based method.}
\label{fig:bottom_line}
\end{figure}

We run hierarchical searches, parameterized by the number of chunks $N_c$ in the 
first level of the hierarchy.  Since the first level dominates the computational cost,
the cost is the same as $N_c$ coherent searches with timestream length $T_c = T/N_c$.
Since the cost of a coherent search is $\bigoh(T_c^3)$, the hierarchical search is
faster than a full coherent search by a factor $N_c^2$.
In each Monte Carlo simulation, we simulate a pulsar with random parameters, and say
that the hierarchical search detects the pulsar if it converges to the full phase model of the
pulsar at the last stage of the hierarchy.  In Fig.~\ref{fig:bottom_line}, we show
the detection probability of the search as a function of $N_c$ and the signal-to-noise.

We emphasize that the total SNR on the $x$-axis is the total signal-to-noise of the
pulsar summed over all $N_c$ chunks.  The hierarchical search succeeds in finding pulsars
whose signal-to-noise per coherent chunk is very small.  For example, a search with
$r=18$ and $N_c=512$ succeeds 90\% of the time, even though the signal-to-noise per
coherent chunk is $18/\sqrt{512} = 0.8$.

To interpret the results in Fig.~\ref{fig:bottom_line}, we will compare to analytic estimates of
some other search methods.

First, we estimate the threshold SNR for a full coherent search as follows.
We assume that in the absence of any pulsar signal, $\hE$ is an independent unit Gaussian 
for each of the $S \approx 2 \times 10^{13}$ models in the search space.  For the
coherent search to succeed, the SNR of the pulsar should be comparable to the expected
maximum of all $S$ Gaussians.  This criterion gives the following estimate:
\be
\frac{1}{S} \approx \frac{1}{(2\pi)^{1/2}} \int_r^\infty e^{-x^2/2} \, dx = \frac{1}{2} \mbox{erfc} \left( \frac{r}{\sqrt{2}} \right)   \label{eq:erfc}
\ee
which gives $r=7.4$ for $S=2\times 10^{13}$.  Remarkably, from Fig.~\ref{fig:bottom_line}
we see that the 50\% detection threshold for
hierarchical search with $N_c=64$ is $r \approx 9$, which is not much worse!
In this case, the hierarchical search does not lose much optimality, but should be faster
than a full coherent search by a factor $\approx 64^2 = 4096$.

The analysis in the previous paragraph does not account for the timestream being
one of many trial (beam, DM) pairs in a full survey.  In a survey, the detection threshold
must be set higher than 7.4$\sigma$, in order to avoid being flooded with false positives
from the large number of trials.
To obtain an estimate for a realistic full-survey detection threshold, we apply Eq.~(\ref{eq:erfc})
with $S$ multiplied by an additional factor $(N_{\rm sky} N_{\rm DM}) \sim 10^9$.
This gives signal-to-noise threshold $r = 9.8\sigma$.  It follows from Fig.~\ref{fig:bottom_line}
that a hierarchical search with $N_c = 64$ is nearly equivalent to a full coherent search, 
since a pulsar which is bright enough to pass the full-survey SNR threshold will be detected
by the hierarchical search 85\% of the time.
A crucial point here is that the last stage of the hierarchical search is a coherent search,
so that if the hierarchical search succeeds then its reported SNR is the true coherent SNR $\hE$
of the pulsar, which can then be compared to the detection threshold of a full coherent search
to reject false positives.

Finally, we would like to compare the hierarchical search to an alternative chunk-based method,
namely incoherent power spectrum stacking.  In this method, we first divide the timestream into chunks
whose size is determined by the criterion that the the first $D^{-1}$ harmonics of the pulse period
drift by $< 1$ Fourier bin during one chunk.  This gives
$T_c \approx (2\pi \alpha_{\rm max}^{-1} D)^{1/2} \approx 80$ sec, 
or $N_c = 167$ chunks.
We next compute the folded power spectrum $P_{\rm folded}(\omega)$ in each chunk as described in
the introduction.  Finally, we loop over trial frequency and acceleration parameters,
and sum $P_{\rm folded}$ values over chunks to get a search statistic $P_{\rm summed}$.  (We need to loop
over trial accelerations since the pulse frequency can drift by more than one Fourier bin between
the first and last chunk.)

To estimate the signal-to-noise threshold for incoherent power spectrum stacking, we will model each 
per-chunk $P_{\rm folded}(\omega)$ value as a $\chi^2$ random variable with $D^{-1}=10$ degrees of freedom.  
A pulsar with total signal-to-noise $r$ will contribute $r^2/N_c$ to $P_{\rm folded}$.
When we sum $P_{\rm folded}$ values over chunks to get $P_{\rm summed}$, we get a 
$\chi^2$ random variable with $d = N_c D^{-1}=1670$ degrees of freedom, and a pulsar contributes $r^2$
to $P_{\rm summed}$.
We estimate that the number of trial frequencies needed for the incoherent search is 
$N_\omega \sim \omega_{\rm max} T_c / (2\pi)$, and the number of trial accelerations is $N_\alpha \sim N_c$.
This gives a total number of trials $N_{\rm trials} = N_\alpha N_\omega \sim 10^7$.
To estimate the threshold signal-to-noise $r$, we require that the expectation value
$(d^2+r)$ of the $P_{\rm summed}$ statistic which contains the pulsar be greater than the expected maximum of
$N_{\rm trials}$ $\chi^2$ random variables with $d$ degrees of freedom.  This gives the criterion:
\be
\frac{1}{N_{\rm trials}} = \int_{d + r^2}^\infty dx \, \frac{x^{d/2-1} e^{-x/2}}{2^{d/2} \Gamma(d/2)}
\ee
where the integrand on the RHS is the PDF of a $\chi^2$ random variable with $d$ degrees of freedom.
This criterion gives signal-to-noise threshold $r \approx 18.6$ for incoherent power spectrum stacking.
Comparing with Fig.~\ref{fig:bottom_line}, we see that incoherent power spectrum stacking
performs substantially worse than the hierarchical search for the same value of $N_c$.
An additional benefit of the hierarchical search is that $N_c$ is a free parameter which can be adjusted
to trade off optimality for computational cost.

\section{Discussion}

In this paper, we have proposed four new algorithms for pulsar search:
an optimal constant-period search (\S\ref{sec:constp}), a recursive tree
algorithm for coherent constant-acceleration search (\S\ref{sec:tree}), 
a semicoherent search which combines information from 
coherent subsearches while preserving as much phase information as possible 
(\S\ref{sec:semicoherent}), and a hierarchical search which uses a sequence of
semicoherent searches to converge to a coherent search (\S\ref{sec:hierarchical}).

The main result is Fig.~\ref{fig:bottom_line}, where we have simulated the
hierarchical search as a function of the number of coherent chunks $N_c$.
The computational cost scales roughly as $N_c^{-2}$, so the free parameter
$N_c$ may be chosen based on total computing time available.
We have shown that for surprisingly large $N_c$, the semicoherent search
is nearly equivalent to a full coherent search.  Therefore, as an optimization,
it may make sense to use $N_c=32$ or 64 by default, speeding up the search
by a factor 1024 or 4096.  If the search is still too expensive, then larger $N_c$ may
be used.  The search will then be suboptimal, but less so than incoherent
power spectrum stacking.

The semicoherent search has a formal interpretation as the optimal search
over the fuzzy search space consisting of all phase models whose 
acceleration $\ddot\Phi$ is allowed to wander over a narrow range $(\Delta\alpha)$.
Intuitively, the fuzziness of the search space means that the phase model
decorrelates on time scales larger than some coherence time $T_c \propto (\Delta\alpha)^{-1/2}$.
The semicoherent search is fully coherent on timescales $\lsim T_c$.
On longer time scales, it starts to become suboptimal,
but has the important property that the computational cost grows slowly as 
$\bigoh(T)$ for $T \gsim T_c$.  One can think of the semicoherent search as
a way of ``fuzzing out'' the constant-acceleration search on long time scales
in order to speed it up.

An interesting property of the semicoherent search is that it should detect
pulsars which are approximated by, but not strictly modelled by, a constant-acceleration
phase model.  For example, a binary pulsar may have a complex phase model with many
parameters, but as long as its acceleration varies by less than $(\Delta\alpha)$ over the
observation, then the semicoherent search should have the same sensitivity to the binary
pulsar as it has to a constant-acceleration pulsar.
In a hierarchical search, the binary pulsar would show up as a statistically significant
peak in the early stages of the hierarchy, which drops out and loses statistical significance
in later stages, when the coherence time $T_c$ becomes large enough that the
constant-acceleration model is no longer a good fit.
In real data, it will be interesting to flag statistically significant hierarchical
search dropouts for human inspection, as a general way of finding pulsars which are
approximately but not exactly fit by the assumed search model.

Another useful property of the hierarchical search is that if it succeeds in finding
the pulsar, then it finds the full phase model and reports the true phase-coherent
signal-to-noise $\hE$.
This is important for rejecting false positives, since the $\hE$-values for the best 
candidates which are found by the semicoherent search may be compared to the detection
threshold for a full coherent search, even though the coherent search is too expensive
to run over the whole parameter space.
This property may also make the hierarchical search more RFI-robust than incoherent power 
spectrum stacking, since weak unmasked RFI events may contribute statistically
to the power spectrum, but are unlikely to have the correct phasing to mimic a
coherent pulsar signal whose likelihood is narrowly peaked as a function
of phase model parameters, beam, and DM.

In the future, we plan to extend this work in several ways.

Our current implementation is not very well optimized, makes
simplifying assumptions such as fixed duty cycle, and is missing
features such as dedispersion and RFI removal.
Because we do not have an optimized implementation, in this paper
we have only been able to make rough estimates of computational
cost, rather than giving hard timings.
We plan to release a public ``production'' version soon which
will address these shortcomings and be applicable to real data.

In this paper, we have only studied the simple case of a constant-acceleration
search space.  We speculate that the methods in this paper may be more powerful
for more complex search spaces.  To take a concrete example, consider the case
of a low-order polynomial search, say degree 3 or 4.  The cost of a full coherent
search grows as $\bigoh(T^6)$ or $\bigoh(T^{10})$ and would quickly become prohibitive.
However, one should still be able to define a semicoherent search whose cost is
the same as a coherent search up to some coherence time $T_c \ll T$, and grows 
linearly afterwards.
The speedup over a coherent search would be even more dramatic than in the 
constant-acceleration case.
One could even imagine a hierarchical search in which the degree of the polynomial
increases during the hierarchy, and in late stages of the hierarchy the polynomial 
fit is replaced by a more detailed model, such as a many-parameter binary system.

An exciting near-term development in radio astronomy will be the advent
of large close-packed transit interferometers such as CHIME~\cite{Bandura:2014gwa}
and HIRAX~\cite{Newburgh:2016mwi}.
These instruments will have mapping speeds hundreds of times larger than
existing telescopes, due to a combination of reasonably large total collecting
area, and very large numbers of formed beams.
However, with existing algorithms it is difficult to
take full advantage of this enormous mapping speed to search for pulsars, 
since the observing time of each sky location is split into noncontiguous
daily observations, and computational cost is also a major issue.
It will be very interesting to see whether the methods in this paper can
be applied to transit telescopes.

As this paper was nearing completion, we became aware of related work
in the context of gravitational wave interferometers such as LIGO.
A variety of statistics have been proposed which search long timestreams
for quasiperiodic signals by combining information from short coherent searches 
(e.g.~\cite{Brady:1998nj,Krishnan:2004sv,Cutler:2005pn,Pletsch:2009uu,Dergachev:2011pd} 
and references therein).  The method we have proposed is conceptually
similar but the details are very different.  In the future, we plan to compare
these algorithms in more detail.

\section*{Acknowledgements}

We thank Scott Ransom and Ue-Li Pen for discussions.
Research at Perimeter Institute is supported by the Government of Canada
through Industry Canada and by the Province of Ontario through the Ministry of Research \& Innovation.
Some computations were performed on the GPC cluster at the SciNet HPC Consortium.
SciNet is funded by the Canada Foundation for Innovation under the auspices of Compute Canada,
the Government of Ontario, and the University of Toronto.
KMS was supported by an NSERC Discovery Grant and an Ontario Early Researcher Award.

\bibliographystyle{h-physrev}
\bibliography{constant_pdot}

\appendix

\section{``Transposed'' coherent search}
\label{app:spacings}

In this technical appendix, we define the notion of a transposed search.
A transposed search $\hE^\dag$ is defined for any coherent search algorithm $\hE$, 
but not for the semicoherent search $\hH$.
This is because a coherent search statistic $\hE$ is a linear function of the timestream data $t_k$,
whereas $\hH$ is nonlinear.

A coherent search algorithm is applied to a length-$N$ timestream $t_k$, and produces a 3D array
$(\hE t)_{\alpha\omega\Phi}$ on a grid of trial accelerations, frequencies, and phases.  
Since $\hE$ is linear, we can view it as a linear operator from an $N$-component
vector space to an $(N_\alpha N_\omega N_\Phi)$-component vector space.

Given two timestreams $t_k$ and $t'_k$, where $k=1,\cdots,N$, we define their dot product by
\be
t\cdot t' = \eta^{-2} t_s \sum_{k=1}^N t_k t'_k
\ee
This agrees with our previous definition in Eq.~(\ref{eq:dot_def}).
Similarly, given two shape-$(N_\alpha,N_\omega,N_\Phi)$ arrays 
$X_{\alpha\omega\phi}$, $X'_{\alpha\omega\phi}$, we define their dot product by:
\be
X \cdot X' = \sum_{\alpha=1}^{N_\alpha} \sum_{\omega=1}^{N_\omega} \sum_{\phi=1}^{N_\phi} X_{\alpha\omega\phi} X'_{\alpha\omega\phi}
\ee
Given these dot product definitions, the transposed search $\hE^\dag$ is defined to be the formal transpose 
of the operator $\hE$ in the usual linear algebra sense.  More precisely, given a 3D array $X_{\alpha\omega\phi}$, 
$(\hE^\dag X)$ is a length-$N$ timestream satisfying:
\be
(\hE^\dag X) \cdot t = X \cdot (\hE t)  \label{eq:Edag_def}
\ee
for all timestreams $t_k$. 
This equation uniquely determines the operator $\hE^\dag$ and can be taken as its definition.

The fast coherent search algorithms from~\S\ref{sec:constp},~\S\ref{sec:tree} are algorithms
which compute $(\hE t)$ for a fixed timestream $t$.  In this appendix we will show that these
algorithms can be formally transposed, to give algorithms for computing $(\hE^\dag X)$ for a
fixed input array $X$.  First we explain why the transposed search $\hE^\dag$ is useful.

Suppose we apply the transpose search operator $\hE^\dag$ to a ``singleton'' array $X_*$, that is a
3D array whose entries are all zero, except for a single entry which is equal to 1.
Let $(\alpha_*, \omega_*, \Phi_*)$ be the trial parameters corresponding to the nonzero entry.  
Now consider Eq.~(\ref{eq:Edag_def}) defining $\hE^\dag$, specialized to the case 
where $X=X_*$ is a singleton array.  The quantity $(X_* \cdot (\hE t))$ appearing on the
RHS is simply the coherent statistic $\hE t$ evaluated at trial parameters $(\alpha_*, \omega_*, \Phi_*)$.
Thus we can write:
\be
(\hE^\dag X_*) \cdot t = (\hE t)_{\alpha_*, \omega_*, \Phi_*}
\ee
On the other hand, from the definition of $\hE$ in Eq.~(\ref{eq:Edef}), the RHS is also equal 
to $(t \cdot \bar I_{\alpha_*,\omega_*,\Phi_*})$, the dot product of $t$ with the normalized
signal timestream $\bar I_{\alpha_*,\omega_*,\Phi_*}$ corresponding to pulsar parameters 
$(\alpha_*, \omega_*, \Phi_*)$.  Therefore, $(\hE^\dag X_*) \cdot t = \bar I_{\alpha_*,\omega_*,\Phi_*} \cdot t$.
Since this applies to any timestream $t$, we must have
\be
\hE^\dag X_* = \bar I_{\alpha_*,\omega_*,\Phi_*}
\ee
We have now derived a key property of $\hE^\dag$.
When $\hE^\dag$ is applied to a singleton array $X_*$, 
the quantity $(\hE^\dag X_*)$ is the normalized signal 
timestream $\bar I_{\alpha_*,\omega_*,\Phi_*}$ associated pulsar parameters
$(\alpha_*,\omega_*,\Phi_*)$ of the singleton array.

\begin{figure}
\centerline{\includegraphics[width=14cm]{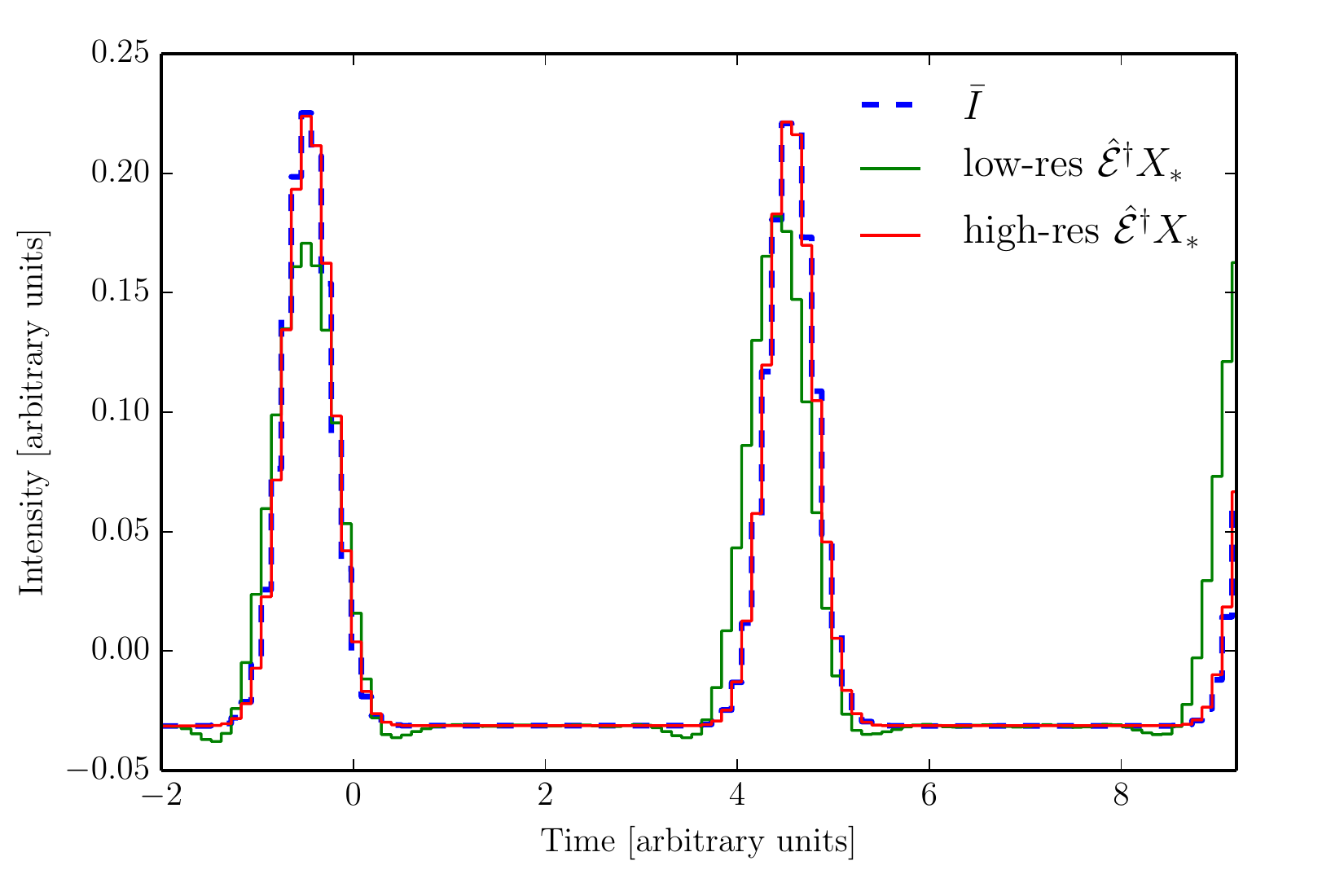}}
\caption{Visual comparison between the transposed search result $\hE^\dag X_*$
for a two-level tree search, and a direct simulation of the pulsar timestream 
$\bar I$ in the time domain, for arbitrarily chosen trial parameters 
$(\alpha_*, \omega_*, \Phi_*)$.
The jaggedness of the curves is due to the discrete timestream sampling.
The two $\hE^\dag$ curves represent choices of resolution-like parameters, such
as spacings in interpolation tables (a complete list of parameters is given in the text).
As the resolution increases, $\hE^\dag$ converges to $\bar I$ precisely, including
discrete-sampling artifacts.
As explained in the text, this gives a complete test that the tree search is
implementing the optimal estimator, and gives a criterion for choosing resolution
parameters.}
\label{fig:visual_test}
\end{figure}

We have found this property to be of great practical use when testing our fast coherent search code.
First, we check that our implementations of $t \rightarrow \hE t$ and $X \rightarrow \hE^\dag X$ are
consistent, by checking that Eq.~(\ref{eq:Edag_def}) is satisfied for randomly generated pairs $(X,t)$.
Then we choose pulsar parameters $(\alpha_*, \omega_*, \Phi_*)$ and verify that $\hE^\dag X_*$ agrees
with a direct simulation of the pulsar in the time domain.
This comparison is shown visually in Fig.~\ref{fig:visual_test}.
Taken together, these tests give a complete test that the coherent search $t \rightarrow \hE t$
is correctly implemented and optimal.
We use this to test the constant-period search algorithm from~\S\ref{sec:constp}
and the constant-acceleration tree search from \S\ref{sec:tree}.

Note that $\hE^\dag X_*$ only agrees with the simulated pulsar timestream $\bar I$ in the limit where
all resolution-like parameters, for example the spacings of trial parameter grids throughout the tree
recursion, are chosen to be high.
Away from the high-resolution limit, there is a difference: $\bar I$ is the true pulsar timestream
which appears in the optimal estimator $\hE_{\rm opt} = (t \cdot \bar I)$, and $\hE^\dag X_*$ is the 
effective pulsar timestream which replaces it in the estimator $\hE$ due to finite resolution artifacts.
The difference between $\bar I$ and $\hE^\dag X_*$ quantifies the suboptimality
due to finite resolution.
Therefore, another use for the transposed search is that it gives us a formal criterion for choosing
resolution-like parameters.
We simply increase resolution until $\hE^\dag X_*$ converges to $\bar I$, for a few random choices of $X_*$.
Here is a complete list of resolution-like parameters, and default settings that we obtained using
this procedure:
\begin{itemize}
\item The spacing $\Delta\alpha$ between trial $\alpha$ values.
  We find that $\Delta\alpha = 100 (D/T_c^2)$ is 3\% suboptimal, and $\Delta\alpha = 70 (D/T_c^2)$ is 0.5\% suboptimal.
  Here, $T_c$ is the timestream chunk size at the current tree resolution.
\item The spacing $\Delta\omega$ between trial $\omega$ values.
  We find that $\Delta\omega = 14 (D/T_c)$ is 4\% suboptimal, and $\Delta\omega = 10 (D/T_c)$ is 0.7\% suboptimal.
\item The number of trial $\phi$ values $N_\Phi$.
  We find that $N_\Phi = (3/2) D^{-1}$ is 3\% suboptimal, and $N_\Phi = 2 D^{-1}$ is 0.7\% suboptimal.
\item The threshold timestream size $T_0$ for switching to a constant-period search, ending the recursion in the tree algorithm.
  We find that $T_0 = 5 (D/|\alpha|_{\rm max})^{1/2}$ is 4\% suboptimal, and $T_0 = 3 (D/|\alpha_{\rm max})^{1/2}$ is 0.6\% suboptimal.
  Here, $|\alpha|_{\rm max}$ is the maximum value of $|\alpha|$ in the search.
\item The amount of zero-padding used in the bottom-level timestream FFT (Eq.~(\ref{eq:domega_def})).
  We find that zero-padding by a factor of two is only 0.1\% suboptimal.
\end{itemize}
In all cases except the last,
we have given one value which is a few percent suboptimal, and a more conservative
value which is $< 1\%$ suboptimal.
We have used the more conservative values as defaults throughout the paper.
When we set all parameters to their defaults, we find (by comparing $\hE^\dag X_*$ to $\bar I$) that
the total suboptimality is 6\%.  This includes the cumulative effect of multiple interpolations in the tree algorithm.

To conclude this appendix, we explain our algorithm for computing $(\hE^\dag X)$
from a 3D array $X_{\alpha\omega\Phi}$.

Let us start with the simplest case, namely the constant-period search from~\S\ref{sec:constp}.
Recall our algorithm for computing $\hE$.
First, we zero-pad the timestream $t_k$ and take its FFT, to obtain the Fourier
transform $\tilde t_j$ at frequencies $\omega = j \omega_f$, where $\omega_f = 2\pi/T_{\rm padded}$ 
is the fundamental frequency of the padded timestream.
Let us write this step as:
\be
\tilde t_j = t_s e^{ij\omega_f (t_s/2-T/2)} \sum_k e^{ij\omega_f k t_s} t_k   \label{eq:app_E1}
\ee
where we have included the prefactor $t_s e^{ij\omega_f (t_s/2-T/2)}$ 
for consistency with the previous definition in Eq.~(\ref{eq:domega_def}).
Second, we interpolate $\tilde t$ to arbitrary frequencies $\omega$.
We write this step as:
\be
\tilde t(\omega) = \sum_j W_j(\omega) \, \tilde t_j   \label{eq:app_E2}
\ee
where $W_j(\omega)$ is a sparse interpolation kernel (we have used cubic interpolation
throughout this paper).
Third, we compute $\hE$ using Eq.~(\ref{eq:constp_fast}), which we repeat here in index notation:
\be
(\hE t)_{\omega \phi} = \frac{1}{A(\omega)^{1/2} \eta^2} \sum_n \rho_n j_0\!\left( \frac{n\omega t_s}{2} \right) e^{in\phi} \, \tilde t(n\omega)  \label{eq:app_E3}
\ee
Given the chain of steps~(\ref{eq:app_E1})--(\ref{eq:app_E3}) defining $\hE$, how do we compute $\hE^\dag$?
We take the defining equation $(\hE^\dag X) \cdot t = X \cdot (\hE t)$ and plug in the above equations for $\hE t$.
\ba
(\hE^\dag X) \cdot t &=& X \cdot (\hE t) \nn \\
  &=& \sum_{\omega\phi} X_{\omega\phi} \frac{1}{A(\omega)^{1/2} \eta^2} \sum_n \rho_n j_0\!\left( \frac{n\omega t_s}{2} \right) e^{in\phi}
        \sum_j W_j(n\omega) t_s e^{ij\omega_f (t_s/2-T/2)} \sum_k e^{ij\omega_f k t_s} t_k \nn \\
  &=& \eta^{-2} t_s \sum_k t_k \left[ \sum_j e^{ij\omega_f k t_s} e^{ij\omega_f (t_s/2-T/2)} \sum_{\omega n} W_j(n\omega) 
        \frac{1}{A(\omega)^{1/2}} j_0\!\left( \frac{n\omega t_s}{2} \right) \sum_\phi e^{in\phi} X_{\omega\phi} \right]
\ea
where in the second line we have plugged in Eqs.~(\ref{eq:app_E1})--(\ref{eq:app_E3}), and in the third line we have rearranged.
In this form, we can read off a formula for $(\hE^\dag X)_k$: it is the expression in brackets in the last line.
Therefore, $(\hE^\dag X)_k$ can be computed from $X_{\omega\phi}$ by the following steps.
\ba
\tilde X_{\omega n} &=& \frac{1}{A(\omega)^{1/2}} \rho_n j_0\!\left( \frac{n\omega t_s}{2} \right) \sum_\phi X_{\omega\phi} e^{in\phi} \nn  \label{eq:app_ET1} \\
\tilde E_j &=& \sum_{\omega n} W_j(n \omega) X_{\omega n} \nn  \label{eq:app_ET2}  \\
(\hE^\dag X)_k &=& \sum_j e^{ij\omega_f k t_s} e^{ij\omega_f (t_s/2-T/2)} \tilde E_j  \label{eq:app_ET3}
\ea
This sequence of steps gives a fast algorithm for computing $(\hE^\dag X)$, with the same computational cost as
our fast algorithm for computing $(\hE t)$.

In fact, one can see that the steps~(\ref{eq:app_ET1})--(\ref{eq:app_ET3}) in the algorithm
for computing $\hE^\dag$ are simply the formal transposes of the steps~(\ref{eq:app_E1})--(\ref{eq:app_E3}),
in the reverse order.
This makes sense, because if a linear operator factors as $\hE = \hE_1 \hE_2 \hE_3$, then its
transpose also factors as $\hE^\dag = \hE_3^\dag \hE_2^\dag \hE_1^\dag$.
Given code for computing $t \rightarrow \hE t$, one can write code for computing $X \rightarrow (\hE^\dag X)$
by a fairly mechanical process, by reversing the sequence of steps and replacing each step by its formal transpose.
Note that the formal transpose of an FFT is another FFT (e.g.~Eqs.~(\ref{eq:app_E1}),~(\ref{eq:app_ET3}) are transposes).
The formal transpose of an interpolation operation, which converts regularly spaced samples of a function to
irregularly spaced samples, is a gridding operation which converts irregularly spaced samples to regularly 
spaced ones (e.g.~Eqs.~(\ref{eq:app_E2}),~(\ref{eq:app_ET2}) are transposes).

This concludes our algorithm for computing $X \rightarrow \hE^\dag X$,
in the case where $\hE$ is the fast constant-period search from \S\ref{sec:constp}.
We now consider the case of the constant-acceleration tree search from \S\ref{sec:tree}.
We outline the steps since the ideas are similar to the constant-period case.

If the depth of the recursive tree search is $d$, then $\hE$ factors as $\hE = \hE_d \cdots \hE_1 \hE_0$, 
where the operator $\hE_0$ is $2^d$ copies of the constant-period search, and $\hE_i$ is $2^{d-i}$ copies 
of a tree merge operator $\hT$, which merges two arrays of shape $(N^{\rm in}_\alpha, N^{\rm in}_\omega, N_\Phi)$
to a single array of shape $(N^{\rm out}_\alpha, N^{\rm out}_\omega, N_\phi)$.  
Since $\hE^\dag = \hE_0^\dag \hE_1^\dag \cdots \hE_d^\dag$, and we have already shown how to compute $\hE_0^\dag$,
it suffices to compute $\hT^\dag$.

Since $\hT$ is an interpolation operator, its transpose $\hT^\dag$ is just the gridding operator
with the same weights.  To write this out in more detail, let us represent $\hT$ as an operator which
operates on an array $X_{abcs}$, where 
$1 \le \alpha \le N_\alpha^{\rm in}$, $1 \le \omega \le N_\omega^{\rm in}$,
$1 \le \phi \le N_\Phi^{\rm in}$, and the index $s$ is $\pm 1$.
Its output is an array $(\hT X)_{\alpha\omega\phi}$, where 
$1 \le \alpha \le N_\alpha^{\rm out}$, $1 \le \omega \le N_\omega^{\rm out}$,
and $1 \le \phi \le N_\Phi^{\rm out}$.
Note that we represent ``input'' indices with roman letters, and ``output'' indices with greek letters.
From Eq.~(\ref{eq:tree}), the tree interpolation $\hT$ can be written:
\be
(\hT X)_{\alpha\omega\phi} = \frac{1}{\sqrt{2}} \sum_{abcs} W_a(\alpha) \, W_b\!\left(\omega + \frac{s\alpha T}{4} \right) W_c\!\left(\phi + \frac{s\omega T}{4} \right) X_{abcs}
\ee
where $W_a, W_b, W_c$ are interpolation weights.  Therefore, the transpose gridding operation $\hT^\dag$ is:
\be
(\hT^\dag X)_{abcs} = \frac{1}{\sqrt{2}} \sum_{\alpha\omega\phi} W_a(\alpha) \, W_b\!\left(\omega + \frac{s\alpha T}{4} \right) W_c\!\left(\phi + \frac{s\omega T}{4} \right) X_{\alpha\beta\gamma}
\ee

\end{document}